\documentclass[conference]{IEEEtran}
\pdfoutput=1
\IEEEoverridecommandlockouts
\usepackage{cite}
\usepackage{amsmath,amssymb,amsfonts}
\usepackage{algorithmic}
\usepackage{graphicx}
\usepackage{textcomp}
\usepackage{rotating}
\usepackage{array}
\usepackage{xcolor}
\usepackage{tikz}
\usepackage{makecell}
\usepackage{multirow}
\usepackage{subfig}
\usepackage{listings}
\usepackage{etoolbox}
\usepackage{dblfloatfix}
\usepackage{changepage}
\def\BibTeX{{\rm B\kern-.05em{\sc i\kern-.025em b}\kern-.08em
    T\kern-.1667em\lower.7ex\hbox{E}\kern-.125emX}}
\begin{document}

\newcommand*\circled[1]{\tikz[baseline=(char.base)]{
		\node[shape=circle,draw,inner sep=0.8pt] (char) {#1};}}

\title{Ensemble Models for Neural Source Code Summarization of Subroutines
	\thanks{This work is supported in part by \emph{redacted for blind review}.}
}

\author{\IEEEauthorblockN{Alexander LeClair, Aakash Bansal, and Collin McMillan}
\IEEEauthorblockA{\textit{Dept. of Computer Science and Engineering} \\
	\textit{University of Notre Dame}\\
	Notre Dame, IN, USA \\
	\{aleclair, abansal1, cmc\}@nd.edu
}}

% \author{\IEEEauthorblockN{\emph{}}
% 	\IEEEauthorblockA{\textit{~} \\
% 		\textit{~}\\
% 		~ \\
% 		~
% }}

\maketitle

\begin{abstract}
A source code summary of a subroutine is a brief description of that subroutine.  Summaries underpin a majority of documentation consumed by programmers, such as the method summaries in JavaDocs.  Source code summarization is the task of writing these summaries.  At present, most state-of-the-art approaches for code summarization are neural network-based solutions akin to seq2seq, graph2seq, and other encoder-decoder architectures.  The input to the encoder is source code, while the decoder helps predict the natural language summary.  While these models tend to be similar in structure, evidence is emerging that different models make different contributions to prediction quality -- differences in model performance are orthogonal and complementary rather than uniform over the entire dataset.  In this paper, we explore the orthogonal nature of different neural code summarization approaches and propose ensemble models to exploit this orthogonality for better overall performance.  We demonstrate that a simple ensemble strategy boosts performance by up to 14.8\%, and provide an explanation for this boost. The takeaway from this work is that a relatively small change to the inference procedure in most neural code summarization techniques leads to outsized improvements in prediction quality.
\end{abstract}


\section{Introduction}

A source code ``summary'' of a subroutine is a natural language description of that subroutine.  Summaries are the foundation of many documentation systems such as JavaDocs, where the summary is used to help programmers quickly gain an understanding of the purpose of the subroutine, without actually reading its source code~\cite{kramer1999api}.  The task of writing these summaries has come to be known as source code summarization~\cite{haiduc2010use}, and has been an active research area for decades because of a mismatch between programmer behavior and their expectations: Programmers tend to avoid writing source code summaries due to the time cost and manual effort~\cite{shi2011empirical, zhong2013detecting}.  Yet at the same time, programmers rely on good documentation written by others~\cite{forward2002relevance}.  The result that automated solutions to code summarization are a very high value target in software engineering research.

Intense research interest has lately been focused on neural source code summarization approaches.  These approaches rely on datasets of millions of examples of code and code summaries (e.g.~\cite{leclair2019recommendations, allamanis2019adverse}) to train a neural network model to predict a source code summary. %  The big data input is possible due to the accomplishments of mining software repositories research, from which we now have paired datasets of source code and summaries of that code with millions of examples and of reliable quality~\cite{leclair2019recommendations, allamanis2019adverse}.  
Almost all neural approaches to code summarization are some form of an attentional encoder-decoder model, in which the encoder creates a vectorized representation of source code, and the decoder represents the natural language summary.  In the past five years, neural approaches have almost completely superseded alternatives such as sentence templates or IR-based keyword extraction~\cite{zhao2020survey, song2019survey, allamanis2018survey}.

Current strategies to neural code summarization can be broadly classified by the type of information they focus on modeling in the encoder.  There are approaches that treat code as text, focusing on the identifier names and other natural language content buried in code~\cite{iyer2016summarizing, loyola2017neural}.  There are approaches that encode the context of other code in the same source file~\cite{haque2020improved}.  Some techniques model the structure of source code with an RNN by linearizing structural representations such as the AST~\cite{loyola2017neural, hu2018deep, leclair2019neural}.  And there is very active scrutiny of GNN-based encoders to model structures such as the AST or CPG~\cite{leclair2020improved, liu2021retrievalaugmented, zugner2021languageagnostic}.  All of these lines of inquiry are showing promise and continue to advance the state of the art.

Recent work in source code summarization has been focusing on providing models with ever more complex representations of code with the assumption that it will yield better and better predictions of code summaries~\cite{leclair2019neural, xu2018graph2seq, hu2018summarizing, hu2018summarizing, haque2020improved}. This approach mirrors progress in many other fields such as machine translation or image captioning~\cite{dabre2020survey, hossain2019comprehensive}.  However, hints from prior work point to a complementary relationship among neural models of code summarization, rather than a competitive one.  Different types of information modeled by the encoder seem to provide orthogonal improvements to the predictions.  For example, LeClair~\emph{et al.}~\cite{leclair2019neural} pointed out at ICSE'19 that their flattened AST approach improved predictions for some Java methods in their dataset but not others -- while they observed an improvement in aggregate BLEU scores, this improvement was concentrated in a subset of methods.  They found a mixture of their approach and a text-only encoder approach to provide the best overall results.  The point is that significant progress may be possible by combining encodings that provide orthogonal improvements.

Meanwhile, ``ensemble'' models have a long tradition in many research areas as a way to combine contributions from diverse data sources~\cite{wang2009diversity, sagi2018ensemble}.  Examples include improving cancer diagnosis accuracy~\cite{hsieh2012design}, weather prediction~\cite{rasp2018neural}, stock market classification~\cite{borovkova2019ensemble}, solar power output efficacy~\cite{al2019ensemble}, and many others.  Sagi~\emph{et al.}~\cite{sagi2018ensemble} point out that the reason ensemble models work is that they increase data source diversity, avoid overfitting, and defray problems related to models stuck at local minima.  Yet the benefits ensemble models may offer to code summarization have not been thoroughly studied.

In this paper, we combine different, complementary source code encoders for summarization via ensemble models.   Our paper has two parts: 1) we conduct an empirical study in which we compare off-the-shelf ensemble technique to combine several baseline code summarization models from related literature, and 2) we provide a rationale behind the increases we observe from the ensemble techniques, showing how different models make different contributions to prediction performance.

% In this paper, we combine different, orthogonal source code encoders for summarization via ensemble models.   Our paper has three parts: 1) we conduct an empirical study in which we use a simple ensemble technique to combine several baseline code summarization models from related literature, 2) we provide a rationale behind the increases we observe from the simple ensemble technique, showing how different models perform better on different subsets of source code, and 3) we use more complex, learned ensemble strategies to achieve state-of-the-art code summarization prediction performance.

We release all code, datasets, and other scripts to help other researchers via our online appendix (see Section~\ref{sec:reproducibility}).

\section{Background \& Related Work}

This section discusses the key supporting technologies and related work to this paper, namely source code summarization in SE research and ensemble models generally.

\vspace{-0.1cm}
\subsection{Source Code Summarization}

% Read paragraph 3 of the introduction.  Note how it broadly classifies code summarization approaches into four categories: text-based, context-based, flat-structure-based, and gnn-structure-based.  Create a table with all neural code summarization approaches you can find.  Categorize them all into one of these four groups.  You can create an ``other'' category if you have to, for things like modifications to the loss/optimization via reinforcement learning or whatever.  Then discuss each group around the table.  This table and discussion should take the rest of this whole page!

Recent work in source code summarization is overwhelmingly centered around the use neural networks and deep learning architectures. The work can be broadly categorized into four research directions:

    \textbf{Text-Based}: Text-Based models use the text of the source code itself as a sequence, relying on the words that appear in the code to generate a summary. These methods usually use a sequence-to-sequence-like model design. The input to these models is the source code being summarized. The intended output is the summary of that source code. These approaches were first described around 2016, such as by Iyer~\emph{et al.}~\cite{iyer2016summarizing}. These approaches are still frequently used as baselines against which to evaluate newer approaches.

  \textbf{Flat-Structure}: Flat-Structure models take the structure of the code, usually in the form of the AST, and flatten it into a sequence. For example, using a depth-first traversal to generate a sequence from the tree. After the early successes of text-based neural approaches, a trend formed in which several papers described code summarization approaches based on this flattened AST structure. For example, Hu~\emph{et al.} create a technique for flattening the AST called structure-based traversal (SBT). Their technique retains the structural information of the AST during the flattening process by adding a series of brackets and braces to group AST nodes. Other, more standard tree traversal techniques such as pre-order or post-order are considered lossy, in that the original AST may not be able to be reconstructed from the flattened output. Alon~\emph{et al.} generated paths between the nodes of the AST as a way to flatten the structure. They randomly selected multiple pairs of nodes in the AST for each method, and use the path between nodes as a flattened representation.

  LeClair~\emph{et al.} observed that the language used in the code and the structure of the code contains orthogonal information and could remain separate inputs to the model. They adapted the SBT approach to what they call the SBT AST Only (SBT-AO), which removed all identifiers from the SBT representation. They then had two inputs, one for the source code sequence and one for the SBT-AO. This allowed the model to learn the information from how the source code is written from one input, while also learning only from the structure with the other input.
  
  \begin{figure}[t]
  \centering
  \scalebox{0.82}{%
    \begin{tabular}{l|c|c|c|c}
 
         &Text&Context&Flat&GNN\\
         \hline
         2016 Iyer \emph{et al.} \cite{iyer2016summarizing}&x&&&\\
         2017 Loyola \emph{et al.} \cite{Loyola2017ANA}&x&&&\\
         2017 Lu \emph{et al.} \cite{lu2017learning}&x&&&\\
         2018 Hu \emph{et al.} \cite{hu2018summarizing}&x&x&&\\
        2018 Liang \emph{et al.} \cite{liang2018automatic}&&&x&\\
        2018 Hu~\emph{et al.} \cite{hu2018deep}&&&x&\\
        2019 LeClair~\emph{et al.} \cite{leclair2019neural}&x&&x&\\
        2019 Alon~\emph{et al.} \cite{alon2018code2seq}&&&x&\\
        2019 Fernandes~\emph{et al.} \cite{fernandes2019structured}&&&&x\\
        2020 LeClair~\emph{et al.} \cite{leclair2020improved}&x&&&x\\
        2020 Haque~\emph{et al.} \cite{haque2020improved}&x&x&x&\\
        2020 Ahmad~\emph{et al.} \cite{ahmad2020transformer}&x&&x&\\
        2021 Z{\"u}gner~\emph{et al.} \cite{zugner2021languageagnostic}&x&x&x&\\
        2021 Liu~\emph{et al.} \cite{liu2021retrievalaugmented}&&x&&x\\
        2021 Bansal~\emph{et al.} \cite{bansal2021projcon}&x&x&x&\\

    \end{tabular}
   }
    \caption{Comparison of recent source code summarization research categorized into four broad categories: Text-Based, Structure-Based, Flat-Structure, and GNN-Structure. These categories reflect the type of input data the models use for source code summarization.}
    \label{tab:scs_compare}
    %\vspace{-0.3cm}
\end{figure}

   \textbf{Context-Based}: Context-Based models rely on information outside of the method or snippet such as API calls~\cite{hu2018summarizing} or other methods in the project~\cite{haque2020improved}. For example, Haque~\emph{et al.}~\cite{haque2020improved} show how other methods from the same file can provide additional needed context for a method summary. In an example, they show a simple setter method ``setIntermediate'' which sets a value to a passed parameter. The comment for this function is ``sets the intermediate value for this flight'' but nowhere in the method does the word ``flight'' appear. Other methods in the file do contain the word flight, since the project in question has to do with getting flight information. Recently, Bansal~\emph{et al}~\cite{bansal2021projcon} developed a project-context method that uses the project context in addition to the method tokens and file context. They created a set of embeddings for each level in the project-file-method hierarchy and provided the model with each representation.
    
    \textbf{GNN-Structure}: GNN-Structure-based models retain structure information in graph or tree formats. For instance, LeClair~\emph{et al.}~\cite{leclair2020improved} build upon their earlier work with flattened structure models by using both a source sequence input and a GNN to learn AST node representations. They found that the AST was able to learn better structure representations than a flattened AST. Liu~\emph{et al.}~\cite{liu2021retrievalaugmented} combine a retrieval based technique and a GNN generated summary to produce summaries. They aggregate summaries from similarly structured code along side a GNN to generate the summary of a method.

\subsection{Ensemble Models}

An ensemble model is one in which several other models are aggregated to generate a single output. Ensemble models are used in a variety of applications ranging from neural machine translation~\cite{garmash2016ensemble} to stock market prediction~\cite{asad2015stock}. Ensemble models have been shown to reach state-of-the-art results in many different areas~\cite{sagi2018ensemble}. The goal of ensembling is to get a ``best of all worlds'' output, in which we can take advantage of a model's relative strengths while simultaneously decreasing the effect of its weaknesses. Ensemble models work by aggregating a collection of models trained for the same task. 

There are a variety of aggregation techniques that are used to combine model outputs. One commonly used aggregation technique is to average the outputs of all the models in the ensemble. When applied to text generation this aggregation technique averages the softmax output of each model for every time step in the sequence. More sophisticated aggregation techniques can also be used, such as an SVM, neural network, or weighted sum. Aggregation techniques that use a learning algorithm are known as meta-learning techniques. By aggregating these different models we can take advantage of orthogonal output. For example, LeClair~\emph{et al.} showed that their non-AST and AST models learned to summarize orthogonal subsets of Java methods, leading them to test a simple ensemble. Their ensemble model outperformed both their non-AST and AST models and showed the potential viability of a more sophisticated ensemble approach.

When working with ensemble models there are three high level design concepts: 1) the data used to train each model, 2) the models used in the ensemble, and 3) the procedure for aggregating those models. When determining model input data, ensemble models are considered either dependent or independent. In a dependent ensemble, each model is dependent on the output from another model. An example of a dependent ensemble model would be AdaBoost~\cite{adaboost}, where subsequent models are trained on previously mis-classified training data. Independent ensembles use a collection of independently trained models. Two of the primary techniques used for independent ensembles is bagging and stacking. In bagging, each model is trained on a subset of the data, while in stacking each model is trained on the entire dataset. In both dependent and independent ensembles the outputs of each model is aggregated to generate a single prediction. In this paper we focus on independent ensemble techniques.

\subsection{Metrics}

 BLEU~\cite{papineni2002bleu} is an automated metric commonly used to score the output of text generation and translation models. BLEU scores are commonly used in source code summarization tasks to rate the summary quality of a model. The BLEU algorithm scores the overlap of N-Grams in a predicted and reference text. To achieve a final single score many researchers use an aggregate of BLEU-1 to BLEU-4 which will count the number of overlaps in 1,2,3,4-Grams of the predicted text in the reference text.

% Now just a tiny subsection on BLEU.  ... Sed nec eros gravida, lobortis purus id, tristique mi. Morbi condimentum enim sed elementum egestas. Nulla suscipit eros facilisis mi interdum ultrices. In et posuere quam. Cras gravida blandit faucibus. Phasellus lacinia dui ante, vitae rutrum felis dignissim sed. Suspendisse consequat vitae ligula non vestibulum. Duis interdum a tellus nec maximus. Suspendisse imperdiet tincidunt elit. Curabitur vel fermentum nunc, convallis pellentesque nisi. Pellentesque quis hendrerit felis. Phasellus laoreet eros tellus, id convallis enim sollicitudin venenatis. Nullam vulputate congue vestibulum. Aenean efficitur molestie.

\section{Research Questions}

The research objective of this paper is to study the effects of ensemble models for neural source code summarization, and to determine the orthogonality of different models that contribute to the ensembles.  We ask the following two research questions:

\vspace{0.1cm}
\begin{description}
	\item[\textbf{RQ$_1$}] What is the performance difference in terms of aggregate BLEU scores of existing baselines, when combined using a simple aggregating procedure?
	\vspace{0.1cm}

	\vspace{0.1cm}
	\item[\textbf{RQ$_2$}] What is the difference in the vector space representations of the functions in the models contributing to the ensembles?
	
	%What is the degree of orthogonality of each model, i.e. over which sets of subroutines does each model deliver the most benefit?
\end{description}
\vspace{0.1cm}

The rationale behind RQ$_1$ is that many models have been proposed for source code summarization, yet related literature does not describe how these models may contribute in an ensemble with other models.  Our goal with this RQ is to cast a wide net and include many different model types, with many different source code input features, while keeping the aggregation procedure simple so that the results may be explained and more easily reproduced. To further focus on the model contribution, we use independent ensemble techniques. We use aggregate BLEU scores because that is by far the most common way in which neural source code summarization techniques are evaluated in related work.

% The rationale behind RQ$_2$ is that while the simple aggregation procedure in RQ$_1$ captures how a simple aggregation procedure performs across a variety of models, there meta-learning approaches that will learn how best to aggregate models. Our aim is to test how meta-learning approaches perform as an aggregation mechanism for model ensembles.

% The rationale behind RQ$_2$ is that while the simple aggregation procedure in RQ$_1$ may capture the overall orthogonality of each model, it does not necessarily capture the best combination of those models. Several ensemble procedures are described in the literature.  Our aim is to select a small set of highly orthogonal models (from RQ$_2$) and determine which of state-of-the-art ensemble procedures leads to the highest quality predictions.

The rationale behind RQ$_2$ is that each model may make orthogonal contributions to the predictions, yet some models may perform best because they provide the most orthogonal view of the source code.  It is useful to know which models make the most different contributions because ensemble models work best by combining diverse inputs~\cite{sagi2018ensemble}. We focus on sets of subroutines for which each model performs the best.

% Lorem ipsum dolor sit amet, consectetur adipiscing elit. Ut a commodo sem, quis tincidunt turpis. Aenean congue, tortor sed dapibus vulputate, dolor nisl maximus ligula, at tincidunt massa ligula et orci. Donec ac augue nisl. Sed fringilla nisi in ante rutrum lacinia. Aliquam sit amet leo tortor. Phasellus et finibus libero. Morbi magna massa, lobortis vel nunc sit amet, vestibulum varius neque. Quisque lacinia efficitur augue, non posuere sapien pretium sed. Duis dictum egestas dui sed tristique. Donec quis odio finibus, sollicitudin diam eget, convallis turpis. Morbi eu massa eu sapien molestie aliquam.

% Quisque maximus rutrum lacus non euismod. In commodo velit purus, eget dignissim lectus accumsan sed. Vestibulum porttitor,

\begin{figure*}[t]
\centering
	\subfloat[t][\label{fig:stacking}]{\includegraphics[width=.5\textwidth]{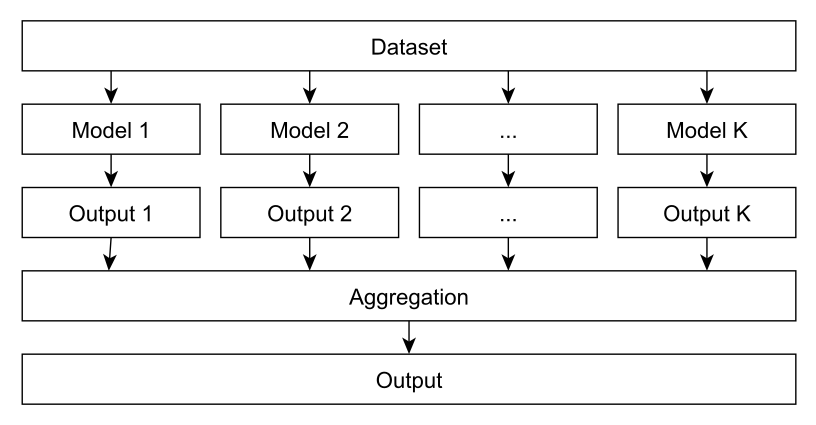}}\hfill
	\subfloat[t][\label{fig:bagging}]{\includegraphics[width=.5\textwidth]{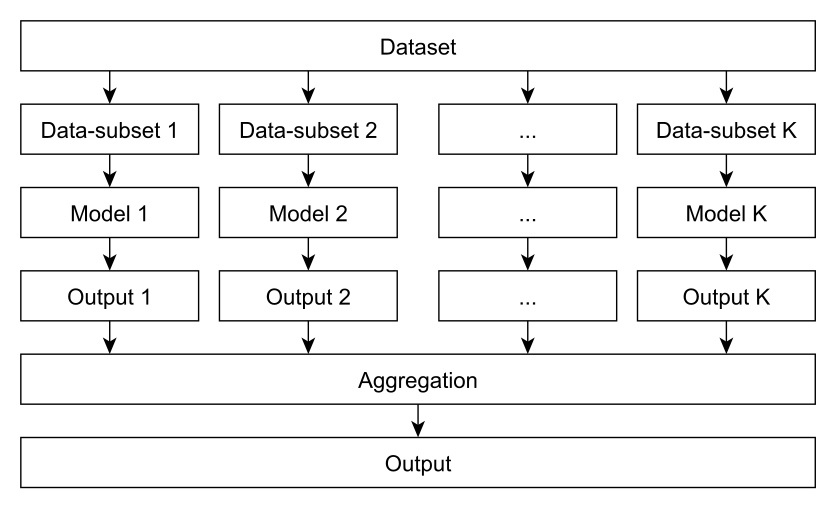}}\hfill
% 	\subfloat[t][Boosting ensemble technique.\label{fig:boosting}]{\includegraphics[width=.333\textwidth]{figures/boosting.pdf}}
    \label{fig:ensemble_diagram}
    \caption{Ensemble architecture diagram for (a) stacking and (b) bagging techniques.}
    %\vspace{-0.3cm}
\end{figure*}

\section{Datasets}

The data used in this work is the Java dataset released by LeClair~\emph{et al.} on recommendations for datasets for source code summarization~\cite{leclair2019recommendations}. This dataset contains ~2.1 million code comment pairs in the Java programming language and is available in two formats, 1) filtered and 2) tokenized. The filtered version of the dataset has taken the raw data and filtered it down to 2.1 million code/comment pairs, but has not applied any additional processing to the text itself. The tokenized format has had text processing applied and is available in a vectorized format. We use the tokenized version of the dataset for this work because it has already been cleaned and processed based on the procedures outlined in other related works~\cite{hu2018deep, leclair2019neural, haque2020improved, alon2018code2seq}. 

We also utilize the adaptation of the Java dataset as outlined in the work by Haque~\emph{et al.}. This adaption adds a set of file context vectors for each method in the dataset. The file context consists of method vectors for each method that exists within the same file. This allows the model to learn additional context from the surrounding methods. We use this set to train the file-context (FC) versions of the models~\cite{haque2020improved}, more details can be found in Section~\ref{sec:baselines}.

\subsection{Threats to Validity}

One threat to validity is the dataset we use. Our dataset contains only a single programming language (Java) and the ensemble models may not generalize to other programming languages. While datasets in other languages are available, few are as large and complete as the Java dataset provided by LeClair~\emph{et al.} and Haque~\emph{et al.}~\cite{leclair2019recommendations, haque2020improved} which contains around 2.1 million examples of source code, comments, ASTs as sequences and trees, and file context vectors. Using this dataset allowed us to use six model architectures from the literature as both baselines and component models for our ensemble techniques.

Ensembling models has a combinatorial explosion affect on the number of hyper-parameters. To limit this effect, we keep many of the hyper-parameters constant across runs and models. We also set our random seed to ensure random selections across model training instances and to reduce the impact of random initialization. Another threat to validity is that we use the BLEU metric to score and compare our models. The goal of this paper is to show how ensemble models perform compared to current literature. Because BLEU score is so common, this allows us to directly compare our models with most other work in this area.

Beyond the combinatorial explosion of tunable parameters introduced by ensembling models, there are also many ways that models can be ensembled and aggregated. For this paper we chose to focus on stacking and bagging to showcase how basic ensemble methods perform compared to individual baseline models. In this paper we constrain our ensemble models to a maximum of two component models. We do this for a variety of reasons. Primarily, in this paper we aim to explore how complementary and orthogonal models contribute to an ensemble, and as a first step to do this we try to reduce the number of tunable parameters. There are also many ways to aggregate models in an ensemble. We chose a simple, easily explainable aggregation method to help reduce the impact that the aggregation method may have on our results.

% Aliquam vel nibh a purus laoreet tempor. Maecenas varius lacus et mi pulvinar imperdiet. Mauris rutrum non nisl ac hendrerit. Duis at volutpat turpis. 

% Nulla volutpat, urna in vehicula gravida, lorem turpis rhoncus dui, gaecenas varius lacus et mi pulvinar imperdiet sed eleifend libero tortor eu libero. Morbi bibendum tortor quis justo lobortis, eget convallis mi tincidunt. Maecenas varius lacus et mi pulvinar imperdiet. Maecenas varius lacus et mi pulvinar imperdiet. Duis sit amet lectus est. Maecenas varius lacus et mi pulvinar imperdiet. Fusce in ligula in sapien consequat ultricies eu eget augue. Aliquam sit amet ligula.
% \begin{figure*}
%     \centering

% \label{tab:results_single}
%     \scalebox{1.0}{%
%     \begin{tabular}{|l|c|c|c|c|c|}
%         \hline
%         Model Type&BA&B1&B2&B3&B4\\\hline
%         Seq2Seq &18.15	&37.87	&20.79	&13.58&	10.15\\
%         Transformer &17.74	&36.81&	20.14&	13.28&	10.07\\
%         Seq2Seq-AST-Flat&19.08&	38.57&	21.78&	14.49&	10.88\\
%         Seq2Seq-AST-GNN&18.74&	37.30&	21.20&	14.33&	10.88\\
%         Seq2Seq-FC&18.15	&36.25	&20.48	&13.85	&10.56\\
%         Seq2Seq-AST-Flat-FC&18.70&	36.92&	21.145&	14.35&	10.92\\
%         \hline
%     \end{tabular}
%     }
% \caption{BLEU 1-4 and aggregate(BA) scores for single model training.}
% \vspace{-0.3cm}
% \end{figure*}

\section{RQ$_1$: Simple Ensembles}
\label{sec:baselines}

This section describes our methodology for answering RQ$_1$ and our results.  Our methodology includes the simple aggregating procedure we used and the baselines we combined.

\subsection{Ensemble Procedure}

For our baseline ensemble methods we use two common ensembling procedures from the literature~\cite{sagi2018ensemble}; stacking and bagging. To reduce the number of variables affecting our results, we restrict our ensemble methods to specific sizes and parameters outlined below.

\vspace{0.2cm}

\begin{itemize}
    
    \item \textbf{Stacking, Figure~\ref{fig:stacking}}: Stacking ensembles aggregate component models that are each trained on the entire training set. Component models in a stacking ensemble do not need to be the same architecture, but they can be. The stacking procedure is 1) select which models will be components in the ensemble, 2) train each model using the entire training set, and 3) select and apply an aggregation method to the output of the models. In our case, each component model in our stacking ensemble is trained on 1.9 million code comment pairs from the Java dataset and a mean aggregation method is applied.
    
    % For source code summarization, this means that during inference we supply our models with the appropriate input based on architecture (source code sequence, AST, file context, etc.) then at each time-step we apply our aggregation method to the softmax output of each model and use the aggregate softmax to select the predicted token from the vocabulary.

    \vspace{0.2cm}

    \item \textbf{Bagging, Figure~\ref{fig:bagging}}: The bagging procedure is very similar to the stacking procedure with one major difference. Instead of training each model on the entire training set, each model is trained on a subset of the training set. There are a variety of ways to choose which subset to train each bagging model on, but a very common method is to randomly select a subset with replacement. In this paper we trained two models on separate randomly selected 50\% of the training set. Because the we randomly select with replacement, there may be overlap in the training data for each model. The intuition behind bagging is that a random subset of data may have a different distribution than the dataset as a whole, allowing the model to learn different sets of features.
    
    % Like stacking, after each model is trained, the chosen aggregation technique is applied to the output of each component model at each timestep, producing a single predicted token.

\end{itemize}

The aggregation technique we use takes the mean of the output at each time step during inference following work in neural machine translation by Sennrich~\emph{et al.}~\cite{sennrich-EtAl:2017:EACLDemo}. In their project, they achieved state of the art translation results by ensembling a collection of trained translation models and using the mean of their output vectors to generate a prediction. With this aggregation technique each model will have slight variations on its output distribution, but we can smooth out these variations by averaging the outputs together.

% For our simple ensemble procedure we use three different ensemble techniques on our models.

% For our simple ensemble procedure we use an average output aggregation function for ever ensemble. The aggregation function takes the set of models, $K=k_1,k_2,...,k_i$, and calculates the mean for each index of the output, $O$, at each timestep, $T$.

% \begin{center}
% {$\forall{t}\in{T}, O_i=\sum{o_i^t}/K$}
% \end{center}

% This aggregation technique has been used in a variety of ensemble approaches such as the neural machine translation system NEMATUS~\cite{sennrich-EtAl:2017:EACLDemo}, which achieved state of the art results for language translation in 2017. While we call this technique `simple' when compared to the other techniques discussed later in the paper, it has been shown to be an effective method for ensembling neural network models~\cite{}.

% We train these ensemble models on a set of four models for Java and two models for C/C++

% % Describe in detail the averaging ensemble technique we use.  Indicate that it is the same one used in Nematus and other tools.  Use mathy symbols and stuff.

\subsection{Baselines}

For our baselines we compare non-ensemble models against stacking and bagging ensemble procedures. We use the baseline models trained following the stacking and boosting procedure as component models in the simple ensemble. We chose these baselines to outline how ensemble methods perform on a variety of model types and architectures, including source code specific features such as the AST and additional file context.

\textbf{Baseline Models}
\begin{itemize}
    \item \textbf{Seq2Seq:} This model is based off the model outlined in Iyer~\emph{et al.}~\cite{iyer2016summarizing}. It uses the source code sequence as input into a standard encoder-decoder architecture. We adapt this model by adding an attention mechanism between the encoder and decoder, which has become common practice for sequence-to-sequence models~\cite{leclair2019neural}.
    
    \item \textbf{Transformer:} We use a transformer encoder baseline following the current trend in neural machine translation~\cite{vaswani2017attention}. Transformers have been shown to outperform sequence-to-sequence models that use recurrent layers in translation and summarization tasks. Because they don't generally use recurrent layers they also train faster than similar GRU or LSTM based models. This baseline uses a transformer to encode the source code sequence and a GRU to decode.
    
    \item \textbf{Seq2Seq-AST-Flat:} This model represents a set of models that use a flattened AST as an input. Hu~\emph{et al.} and LeClair~\emph{et al.} use an SBT representation of the AST, flattening it to a single sequence. LeClair~\emph{et al.} additionally remove identifiers from the SBT representation calling the new representation SBT-AST-ONLY (SBT-AO). We use the SBT-AO representation of the AST for our flat-AST baseline becuase it has been shown to outperform the SBT approach. In this approach both the source code and flattened AST sequence are provided to model as input.
    
    \item \textbf{Seq2Seq-AST-GNN:} This model uses a GNN to encode the AST instead of flattening it. We follow work outlined in LeClair~\emph{et al.}, Fernandes~\emph{et al.}, and Xu~\emph{et al.}~\cite{leclair2020improved, fernandes2019structured, xu2018graph2seq}. In these works, the AST is kept as a tree with each AST node becoming an input to the encoder. A GNN layer is then used to encode the AST nodes with attention mechanisms between the source code, AST, and generated comment.
    
    \item \textbf{Seq2Seq-FC:} The FC version of the seq2seq model follows the work by Haque~\emph{et al.}~\cite{haque2020improved} which uses the other methods from the same file as additional context to the model. This additional file context allows the model to learn vocabulary that may not exist within the method itself. Other related work includes additional context information such as API calls~\cite{hu2018deep} or project level information~\cite{bansal2021projcon}. We chose to use file context because the dataset was readily available and in their work, Haque~\emph{et al.} compares to a variety of additional baselines.
    
    \item \textbf{Seq2Seq-AST-Flat-FC:} Similar to the seq2seq-FC baseline, this baseline uses the file context model proposed by Haque~\emph{et al.}~\cite{haque2020improved} with the addition of the flattened AST as an input. This model has three inputs, the source code sequence, the flattened AST sequence, and the additional file context.

\end{itemize}

\begin{figure*}
\centering
\subfloat[\label{tab:results_single}]{
    \begin{tabular}{|l|c|c|c|c|c|}
        \hline
        Model Type&BA&B1&B2&B3&B4\\\hline
        Seq2Seq             &  18.15            &  37.87   &  20.79            &  13.58            &  10.15\\
        Transformer         &  17.74            &  36.81            &  20.14            &  13.28            &  10.07\\
        AST-Flat    &  19.08   &  38.57            &  21.78   &  14.49   &  10.88\\
        AST-GNN     &  18.81            &  37.70            &  21.33            &  14.33            &  10.85\\
        Seq2Seq-FC          &  18.15            &  36.87	        &  20.52            &  13.73            &  10.44\\
        AST-Flat-FC &  \textbf{19.31}            &  \textbf{38.62}            &  \textbf{21.83}            &  \textbf{14.70}            &  \textbf{11.22}\\
        \hline
    \end{tabular}
}

\subfloat[]{
	\label{tab:results_stacking}
	\scalebox{0.85}{
		\setlength\extrarowheight{10pt} % MIGHT NEED TO TURN THIS BACK OFF LATER
		\begin{tabular}{l|c|c|c|c|c|c}
			\textbf{}& \rotatebox{90}{seq2seq} & \rotatebox{90}{Transformer}& \rotatebox{90}{AST-Flat} & \rotatebox{90}{AST-GNN} & \rotatebox{90}{seq2seq-FC} & \rotatebox{90}{AST-Flat-FC}   \\\hline
			seq2seq     &    19.34 &    19.36 &    19.86 &   19.81  & 19.92 &    20.42    \\\hline
			Transformer & x     & 18.88    &  19.8 &  19.64   & 19.71 &  20.38  \\\hline
			AST-Flat    & x     & x     & 20.13    &  20.13& 20.26 &  \textbf{20.64}   \\\hline
			AST-GNN     & x     & x     & x     & 19.89    & 20.09 &   20.22   \\\hline
			seq2seq-FC  & x     & x     & x     & x     &    19.36   &  20.20    \\\hline
			AST-Flat-FC &  x    & x     & x     &x      & x     &    20.16  \\\hline
			
		\end{tabular}
    }
}
\hspace{0.1cm}
\subfloat[]{
\label{tab:results_bagging}
    \scalebox{0.85}{
		\setlength\extrarowheight{10pt} % MIGHT NEED TO TURN THIS BACK OFF LATER
		\begin{tabular}{l|c|c|c|c|c|c}
			\textbf{}& \rotatebox{90}{seq2seq} & \rotatebox{90}{Transformer}& \rotatebox{90}{AST-Flat} & \rotatebox{90}{AST-GNN} & \rotatebox{90}{seq2seq-FC} & \rotatebox{90}{AST-Flat-FC}   \\\hline
			seq2seq     & 18.53  &  18.43 & 18.89  & 18.92  & 18.97  &   19.28   \\\hline
			Transformer &  x    & 18.07  &  18.88 & 18.86  &  18.82 & 19.06   \\\hline
			AST-Flat    & x     & x     & 19.34  &  18.88 & 19.34  &  \textbf{19.63}    \\\hline
			AST-GNN     & x     &x      & x     & 19.11  & 19.30  &  19.50    \\\hline
			seq2seq-FC  &  x    & x     & x     & x     & 18.01  &  18.98    \\\hline
			AST-Flat-FC &  x    &  x    &  x    & x     & x     &  18.31    \\\hline
		\end{tabular}
	}
}
\caption{BLEU scores for (a) non-ensemble models, (b) stacking with simple aggregation, and (c) bagging with simple aggregation}
\label{fig:results}
%\vspace{-0.2cm}
\end{figure*}

\subsection{Results}

We present our experimental results in three parts for comparison, 1) baseline models evaluated without ensembling (Figure~\ref{tab:results_single}), 2) combinations of baseline models trained using the stacking procedure (Figure~\ref{tab:results_stacking}), and 3) combinations of baseline models trained using the bagging procedure (Figure~\ref{tab:results_bagging}). 

In Figure~\ref{tab:results_single} we present BLEU scores for the set of baseline models without any ensembling procedure applied. We observe that the AST-Flat-FC model had the best overall performance with a BA score of 19.31. The Transformer model had the lowest performance with a BA score of 17.74. The Seq2Seq and Seq2Seq-FC models had the same performance with a BA score of 18.15. The AST-Flat model obtained a 19.08 BA score and the AST-GNN model obtained an 18.81 BA score. We use these baseline scores as a direct comparison for the stacking and bagging ensemble procedures. While the results we obtained for the baseline models is in line with those reported in their respective papers, we do see some variation when comparing BA scores. Difference in score can be attributed to random initialization of parameters and differences in training, validation, and testing set split. 

In Figure~\ref{fig:stacking} we show the results for the stacking procedure using the set of baseline models as component models. The diagonal of Figure~\ref{fig:stacking} show the results for ensembles whose component models are the same architecture. Outside of the main-diagonal are results for each combination of component models using differing model architectures. 

First, if we look only at ensembles whose components are the same architecture (the main diagonal of Fig~\ref{fig:stacking}) we see that the AST-Flat-FC ensemble has the best performance with 20.16 BA score, which is a 4.7\% increase over a single AST-Flat-FC model. The ensemble with lowest performance from this group is the Transformer ensemble with 18.88 BA score. Every model, when having components of the same architecture, achieved an improved BA score with an average increase of 1.10 BA, or an average of 6\% improvement. The Seq2Seq-FC ensemble had the largest increase in performance with a 1.21 score improvement. We attribute the increase in performance when component models use the same architecture to a smoothing effect that combining the outputs has on the prediction. 

Second, when we compare ensembles that use component models of different architectures, we see that the AST-Flat-FC+AST-Flat ensemble achieved a BA score of 20.64. This shows an 8.1\% improvement over the AST-Flat model and a 6.8\% improvement above the AST-Flat-FC model. The worst overall performing ensemble when using two different model architectures is the Seq2Seq+Transformer ensemble. This combination resulted in a BA score of 19.36. The ensemble that had the largest improvement over the baseline model is the Transformer+Seq2Seq-AST-Flat-FC ensemble. This combination achieved a 14.8\% BA improvement over the baseline Transformer model, and a 8.9\% improvement over the baseline AST-Flat-FC model. The stacking procedure resulted in improvement on every ensemble combination. We believe that the improved score is due to the combination of complimentary and orthogonal information provided by the combination of models.

Figure~\ref{tab:results_bagging} shows our results from the bagging procedure using the set of baseline models as component models. When the bagging procedure is applied to two model of the same architecture the AST-Flat ensemble has the best BA score of 19.34, which is a 0.03 BA improvement over the baseline AST-Flat model. The Seq2Seq-FC+AST-Flat-FC ensembles see a performance decrease when trained using the bagging procedure. This performance loss can be attributed to the reduced training data size introduced by bagging. The FC models use additional file context inputs to help improve model performance, but this adds additional trainable parameters to the model. We also note that the FC models need to be trained to 15 epochs before convergence as reported by Haque~\emph{et al.} while the other models only need 10 epochs to converge. It is likely that these models require more data than the other models due to the number of trainable parameters.

When the bagging procedure is applied to component models of different architectures the AST-Flat+AST-Flat-FC ensemble has the best performance with a BA score of 19.63. The Seq2Seq+Transformer ensemble had the worst overall performance with a BA score of 18.43. We found that the combination of models that had the same input types (e.g. the Seq2Seq and Transformer, or AST-Flat and AST-GNN) generally had minimal performance increase from ensembling. Models that have different or orthogonal data inputs had larger BA score improvements when ensembled.

Overall, we found that bagging performed worse than stacking. This could be due to a variety of factors. First, the complexity of the component models may require a large dataset for model convergence, which bagging reduced the training set size significantly. Second, bagging performance may be sensitive to the number of component models. We limit our work to two component models, but more models may improve performance. The stacking procedure had improvements over all baseline component models with the largest improvement being the Transformer+AST-Flat-FC ensemble, and the best combination of component models being the AST-Flat+AST-Flat-FC ensemble. The stacking ensemble results show that models that were trained using complementary orthogonal input data have the best improvement over their baseline models.

\begin{figure}[b]
\vspace{-0.2cm}
\centering
\includegraphics[width=\columnwidth]{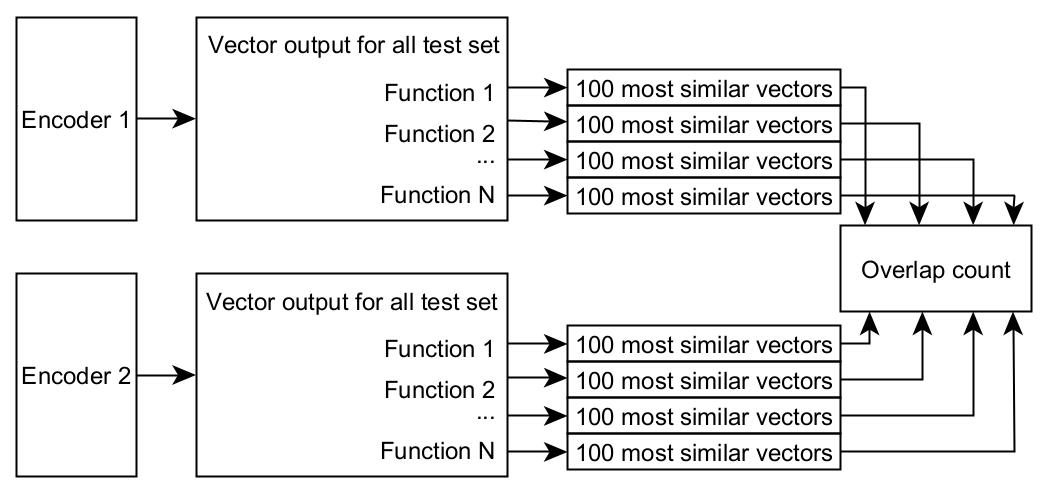}
\caption{Methodology for finding orthogonal representations of source code features.}
\label{fig:orthogonal_methodology}
\vspace{-0.3cm}
\end{figure}

\section{RQ$_2$: Vector Space Analysis}

This section describes our methodology for answering RQ$_2$, procedures for analysis, data, and interpretations.
\vspace{-0.1cm}
\subsection{Methodology}

To answer RQ2 we evaluate and compare how the model encoders create internal representations of the source code. Comparing the output of the model encoders can show us if the models are encoding methods in a similar way. Due to random initialization and weight updates during training, we can not directly compare the output vectors of each model. Instead we compare sets of similar methods using a cosine-similarity score. A high level overview of the process we use to obtain the similar methods can be seen in Figure~\ref{fig:orthogonal_methodology}.
\vspace{0.1cm}

\noindent Procedure to extract similar methods in the testing set using cosine-similarity:

\begin{enumerate}
    \item Using a trained model, extract the vector output of the encoder for every method in the testing set. This gives us the models learned internal representation of each method in the testing set. Do this for both encoders in the comparison. This produces two lists of vector representations for each method.
    \item For the file context vectors, we average the output of the time distributed GRU layer.
    \item For each method vector in the testing set, we find the 100 most similar method vectors using cosine-similarity metric. We apply this to the lists generated by both encoders.
    \item To compare the different encoders, calculate the number of functions that overlap between each methods top 100 similarity list.
\end{enumerate}
\vspace{0.1cm}

Using this method we compare the outputs of model encoders for the source code, AST, and file context of the baseline models. We found that the encoders that have many overlapping functions, then it is likely that the encoders have learned to represent methods in a similar way. If there is low agreement between the encoders list of similar functions, then the encoders may have learned orthogonal representations of the source code. Using our results from RQ1 we show that the models that had the largest improvement when ensembled, also show very little overlap in their inputs when compared with other models.

\begin{figure*}[hbtp]
\vspace{-0.5cm}
\centering
\subfloat[Source code sequence encoder and AST encoder comparison of the AST-Flat model. \label{fig:histogram_astattendgru}]{
\includegraphics[trim=100 70 100 100,width=0.8\columnwidth]{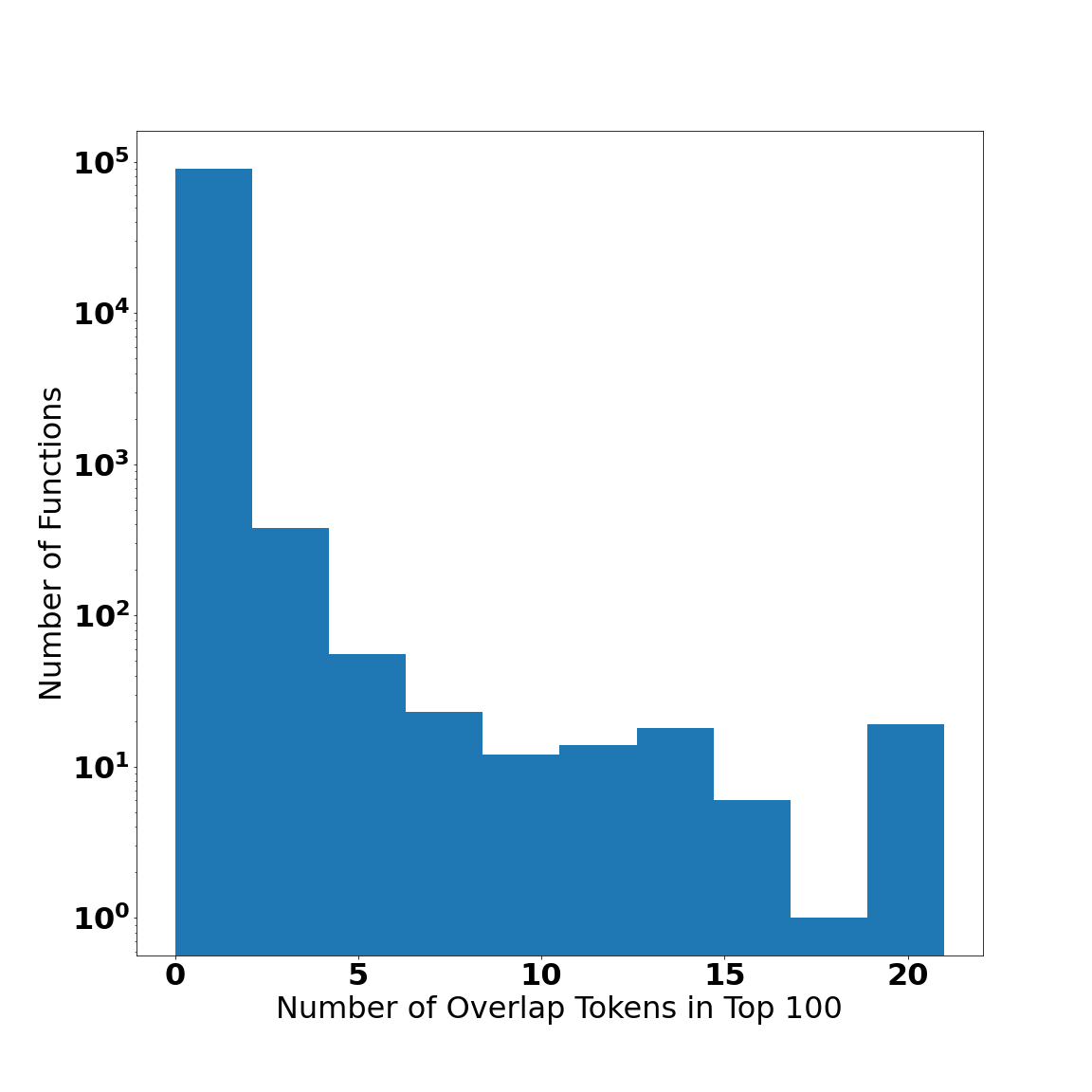}}
\hspace{2cm}
\subfloat[Source code sequence encoder and AST encoder comparison of the AST-GNN model.
\label{fig:histogram_codegnngru}]{
\includegraphics[trim=100 70 100 100,width=0.8\columnwidth]{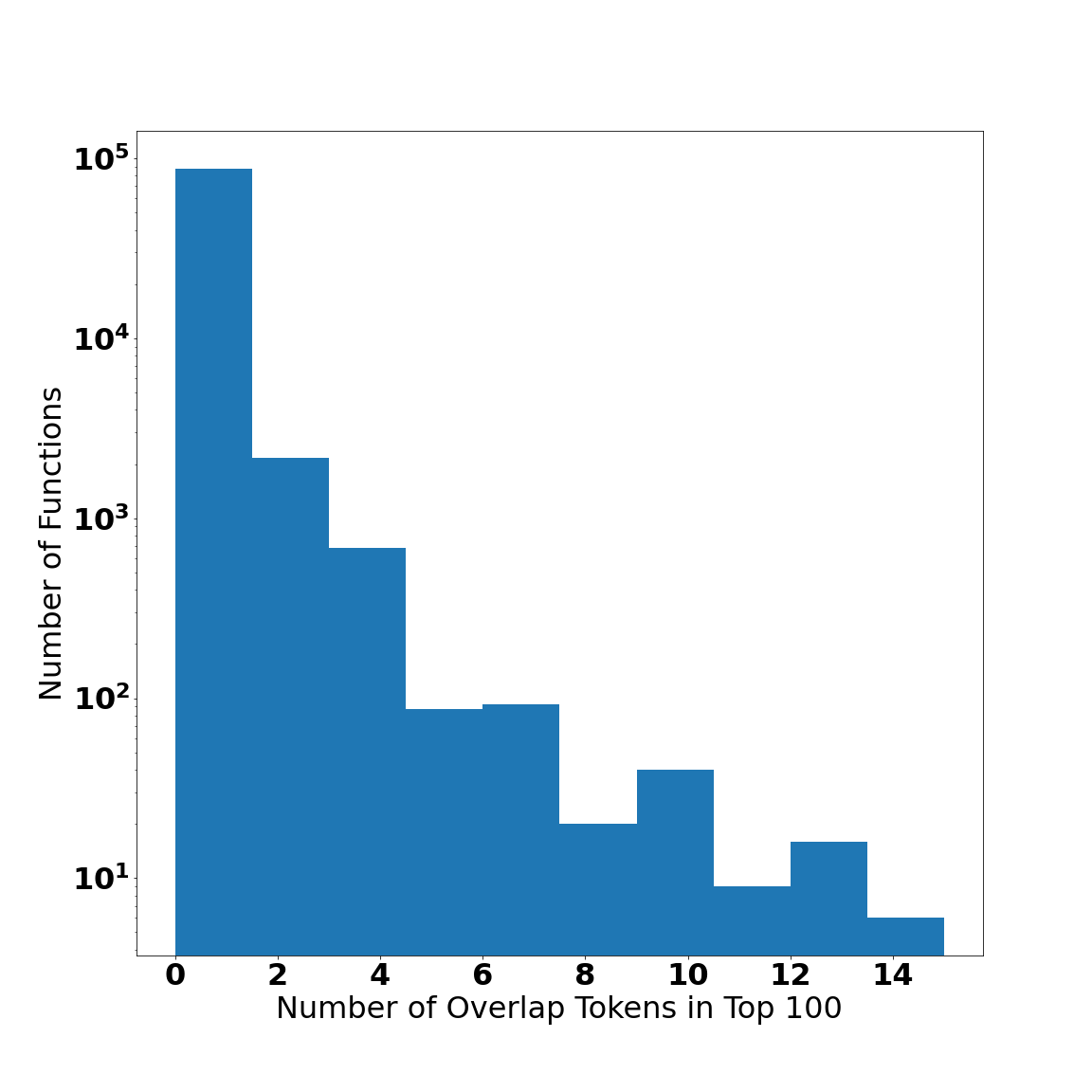}}

\subfloat[AST encoder comparison of the AST-Flat and AST-GNN models.
\label{fig:astattendgru_codegnngru_ast}]{
\includegraphics[trim=100 70 100 100,width=0.85\columnwidth]{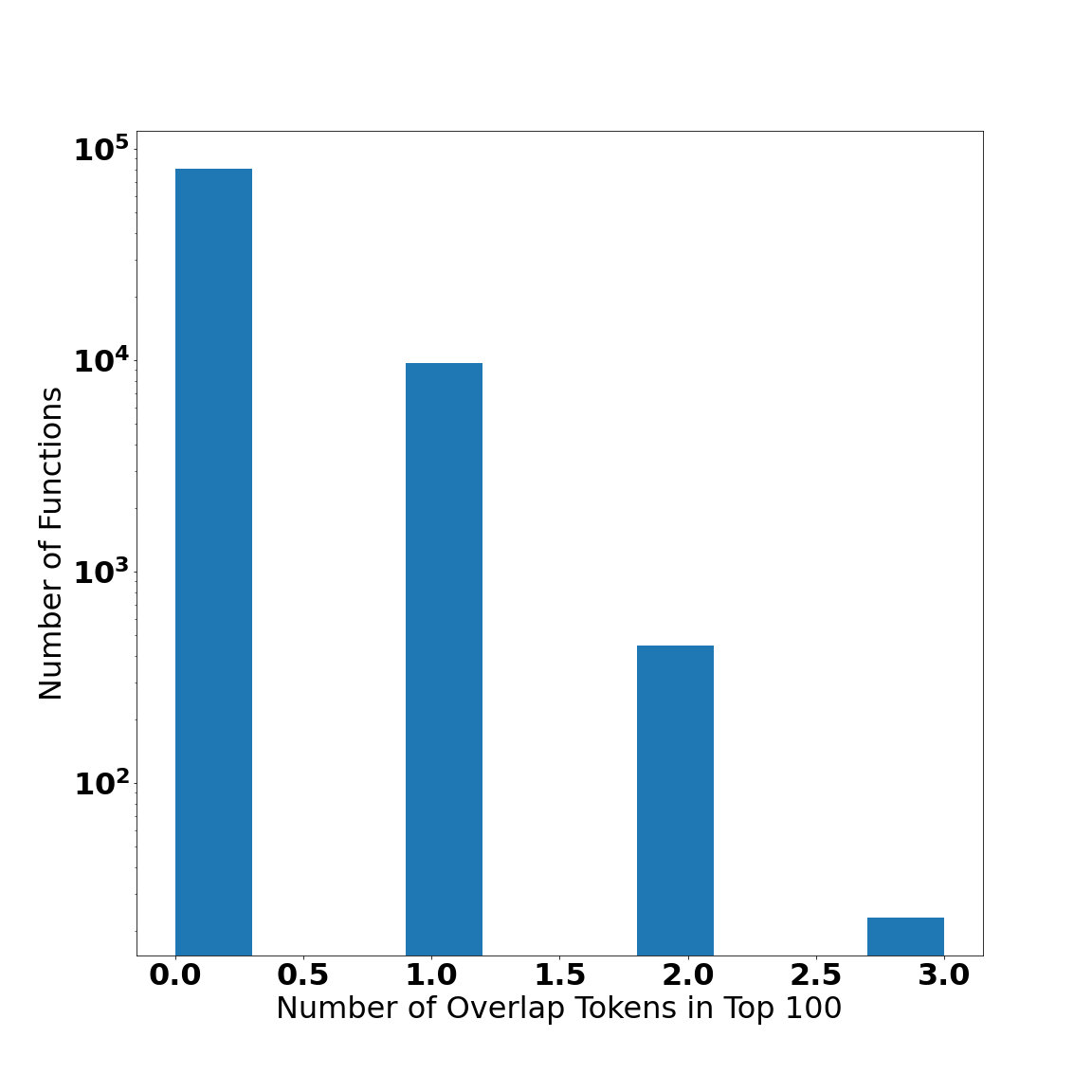}}
\hspace{2cm}
\subfloat[Source code sequence encoder comparison of the Transformer and AST-Flat models.\label{fig:transformer_astattendgru_tdats}]{
\includegraphics[trim=100 70 100 100,width=0.8\columnwidth]{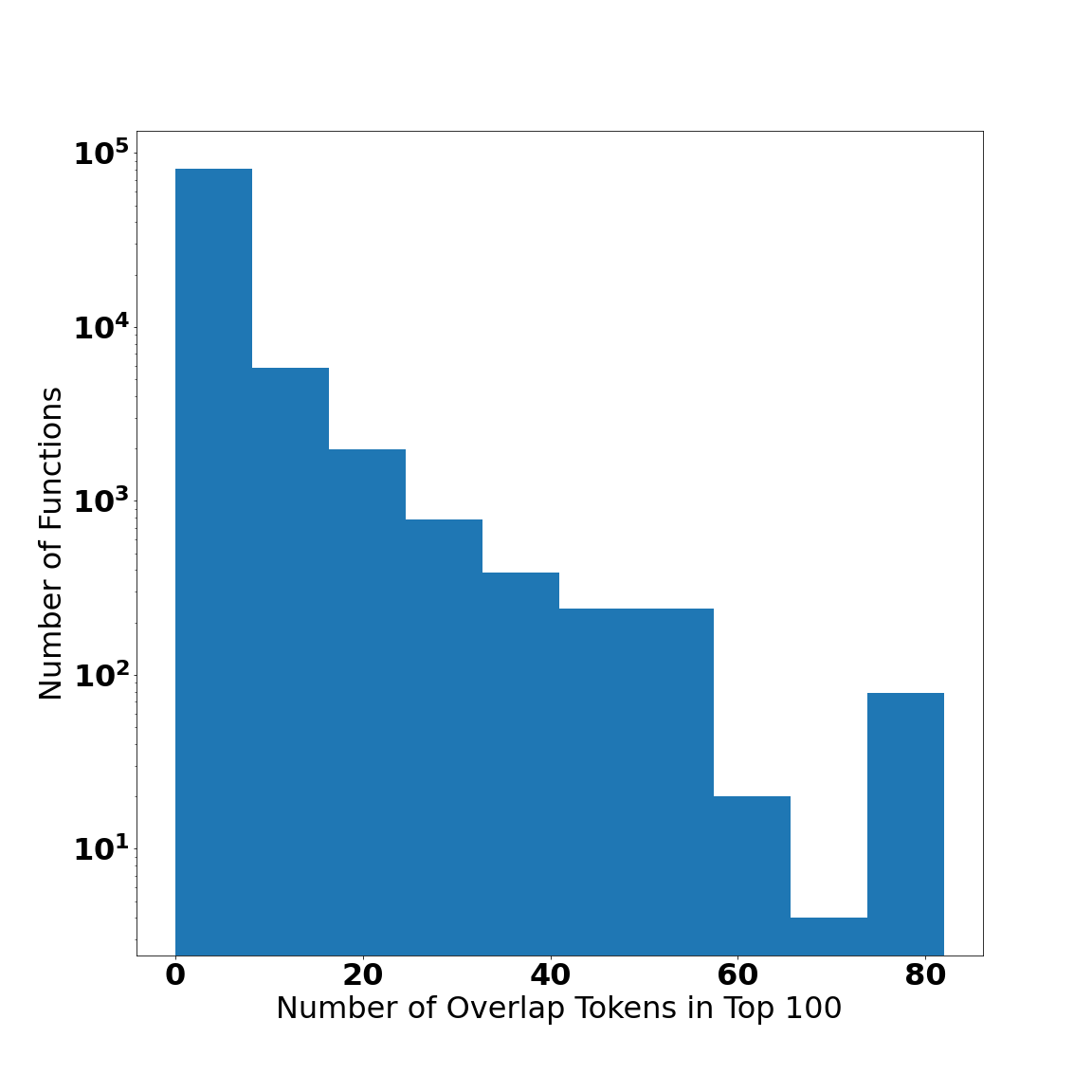}}

\subfloat[File context encoder from the Seq2Seq-FC and AST-Flat-FC models.
\label{fig:astattendgrufc_attendgrufc_fc}]{
\includegraphics[trim=100 70 100 100,width=0.8\columnwidth]{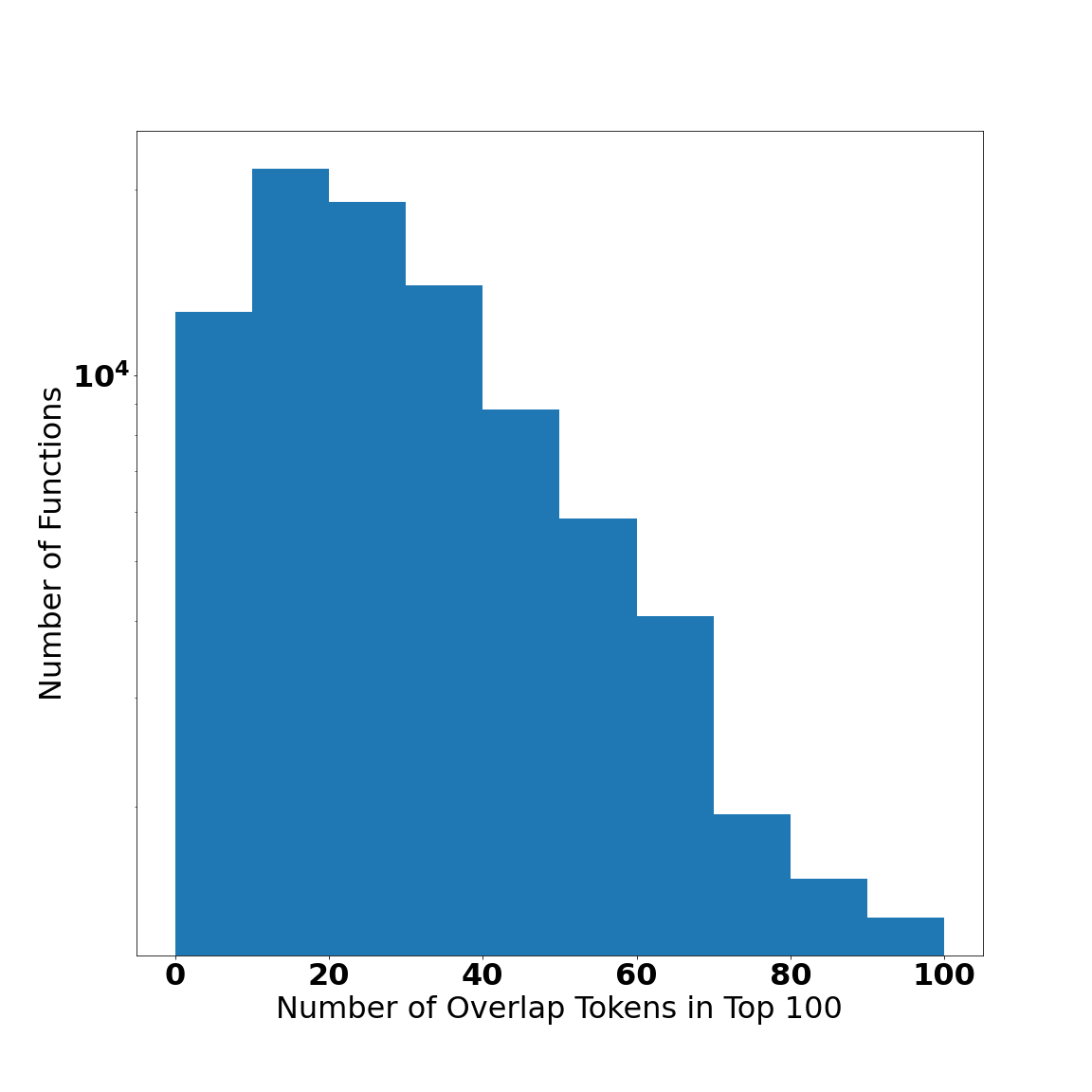}}
\hspace{2cm}
\subfloat[AST encoder for AST-Flat and AST-Flat-FC models.\label{astattendgru_ast_astattendgrufc_ast}]{
\includegraphics[trim=100 70 100 100,width=0.8\columnwidth]{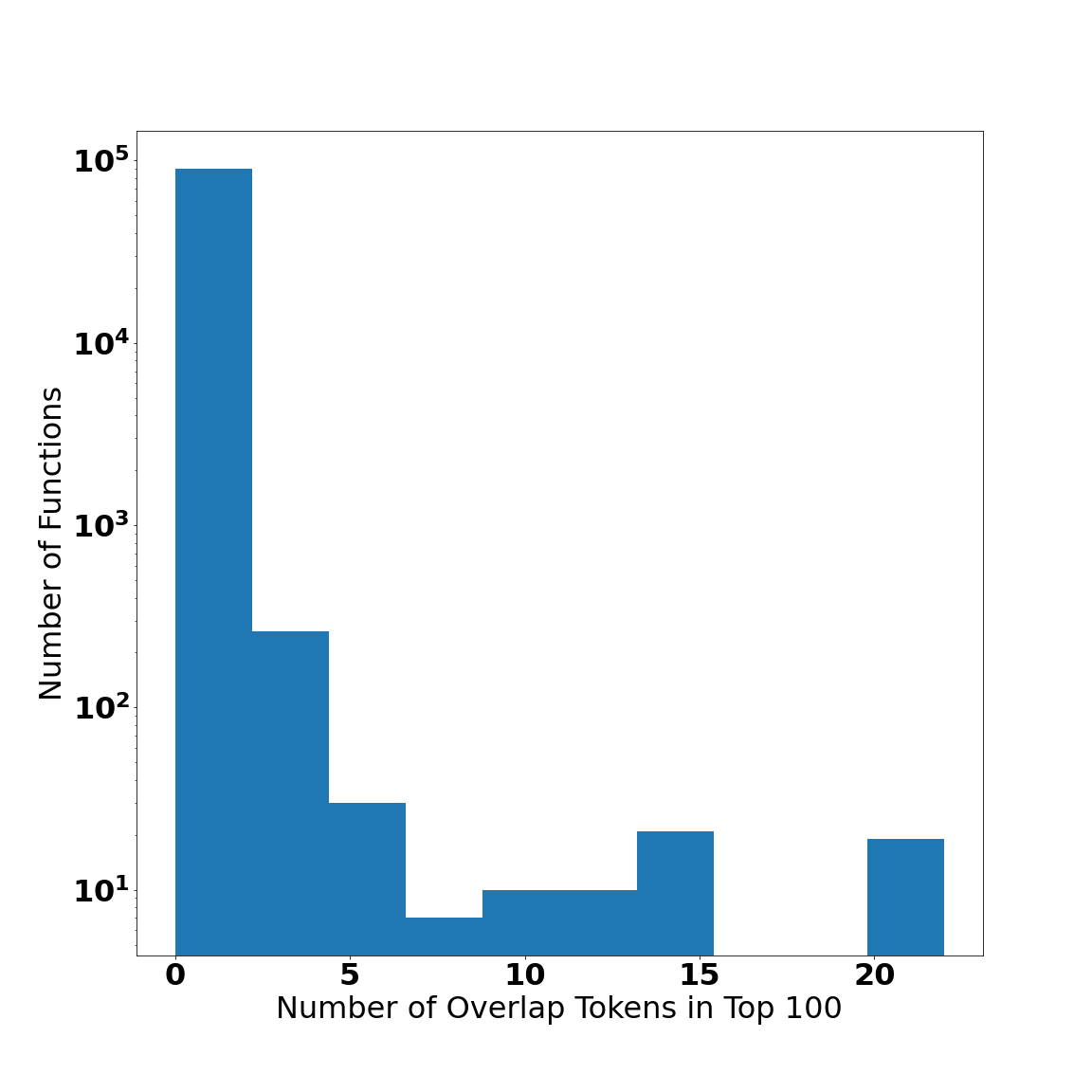}}

\caption{The number of methods in the 100 most similar that overlap between encoders}
\label{fig:histograms}
\end{figure*}

\subsection{Results}

Figure~\ref{fig:histogram_astattendgru} shows histograms of encoder comparisons. In these figures the x-axis is the number of methods that overlap in the encoders 100 most similar lists, as explained in the previous section. For example in Figure~\ref{fig:histogram_astattendgru}, the bar labeled `5' is the count of all methods in the testing set that had an overlap count of 5. Having an overlap count of 5 indicates that for a given method the encoders agreed on 5 entries. In this example, the number of methods that had an overlap count of 5 in the testing set is in between 100 and 1000. Each histogram shows the overlap distribution over the testing set. 

Figure~\ref{fig:histogram_astattendgru} is the overlap between the source code sequence encoder and AST encoder from the AST-Flat model. When comparing the source code sequence encoder to the AST encoder we see very little overlap in method similarity. This may indicate that the source code sequence encoder and the AST encoder have learned to represent methods in different ways. This is likely due to the encoders learning from orthogonal, complementary information. We expect this from inputs that use different parts of the source code, in this case the source code sequence and AST.

In Figure~\ref{fig:histogram_codegnngru} we see a similar situation where there is almost no overlap in the methods the encoders find similar. This figure compares the source code sequence encoder and AST encoder of the AST-GNN model. We see an overlap histogram similar to the AST-Flat comparison, with less overlap in the 10+ groups, but slightly more in the 0-3 groups. This could be due to the GNN providing a more orthogonal representation of the AST than the AST-Flat model. In Figure~\ref{fig:astattendgru_codegnngru_ast} we show the overlap between the AST encoders of the AST-Flat and AST-GNN models. In this histogram we see that while there is still very little overlap, there are many more methods that have an overlap count of 1 or 2. This could mean that while the ASTs are being represented differently to each model, they are learning some features that allow them to identify a small subset of similar methods the same way. 

Not all encoder overlaps show orthogonal representations. In Figure~\ref{fig:transformer_astattendgru_tdats} there is significantly more overlap between the encoders, with some methods having 70-80 of their top 100 similar methods shared between the encoders. This figure shows the overlap between the source code sequence encoders from the Transformer and AST-Flat models. Both encoders are trained on the source code sequence as input and have learned similar representations of the source code sequence input. We found that many of the of the source code sequence encoders shared a high level of similarity. Similarly, in Figure~\ref{fig:astattendgrufc_attendgrufc_fc} we compare the file context encoder average output from the Seq2Seq-FC and AST-Flat-FC models. In this comparison we see a lot of overlap between the encoders, the only difference between these two models is that the AST-Flat-FC model has an additional input of the flattened AST. We attribute the large overlap between these encoders to the file context being learned in similar ways between the Seq2Seq-FC and AST-Flat-FC models. It is likely that these models utilize the file context in similar ways when generating summaries.

Figure~\ref{astattendgru_ast_astattendgrufc_ast} compares the AST encoder of the AST-Flat and AST-Flat-FC models. Again, we see very little overlap between the AST encoders. We found this to be common between all of the models that utilize either flat or GNN representations of the AST. This could be due to each model learning different types of features from the AST that, along with the other inputs, improve model performance. For instance, having the file context along with the AST may allow the model to focus on AST structure elements to boost performance on a specific subset of methods. If the model does not have the file context available, it may have to learn more generalized representations for the AST.

% for teh Seq2Seq-FC and Seq2Seq-AST-Flat-FC models. This shows that these two encoders likely learned to represent methods in a similar way. In this case the encoders are the source sequence encoders for the Seq2Seq-FC and the Seq2Seq-Ast-Flat-FC models. 

% In this figure we see much more overlap between the encoders. The two encoders that we used to generate this histogram are the AST encoder from the AST-Flat model and the AST encoder from the AST-GNN model. Both of these encoders are learning similar representations of the code from the AST, but not exactly the same because of the different AST representation. If we look at Figure~\ref{} we see the AST-GNN encoder compared the the source code text encoder from the same model. We see that this has more overlap than the same comparison in Figure~\ref{} of the AST-Flate source and AST encoder. 

% This gives us a clue that the GNN representation of the AST may be allowing the encoder to learn a representation of the source code that is more similar to how the source code sequence encoder represents the same code. 

% \begin{figure}[b]
%     \centering
%     \includegraphics[width=0.9\columnwidth]{figures/src_text_cluster_astattendgru.png}
%     \caption{Caption}
%     \label{fig:my_label}
% \end{figure}

\section{Examples}
In this section we give two examples from the testing set. These examples show how the encoders for the source code, AST, and file context differ in how they represent functions and determine which functions are most similar.

\subsection{Example 1}

This example compares similar methods from the source code sequence encoder and AST encoder of the AST-Flat model. It shows how the two encoders differ by showing the most similar method to the input method for both encoder. We show the input method source code and source sequence, as well as the associated comment.

The input method is a table GUI method that returns a Boolean if a tooltip is able to be set. The most similar method based on the source code encoder is another GUI method that returns a Boolean and determines if a panel is in a certain position. The most similar method based on the AST encoder is a networking method. This method also returns a Boolean similar to the input method, but it doesn't have much else in common. The AST encoder is finding structural elements to match to. In this case the return type of the method. The source code sequence encoder is matching to similar vocabulary such as `table', `graph', `panel`.

\vspace{0.2cm}
\noindent\textbf{Input Method ID: 18252737}
\vspace{0.4cm}

	\noindent\emph{\textbf{Source code input}} 
	{\small
	\begin{verbatim}
private boolean showTable(Graph graph,...

  if (table!=null&&!tableNodes.containsKey...
  
    Node n = graph.addNode();
    
    n.setString("label", table.getName());
    
    String tooltip = tableRenderer.get...
    n.setString("tooltip", tooltip);
    
    tableNodes.put(table, n);
    return true;
  }
  
  return false;
  
}
	\end{verbatim}}
	
	\noindent\emph{\textbf{Source code sequence}} 
	\begin{adjustwidth}{0.5cm}{0.5cm}
		private boolean show table graph graph table table if table null table nodes contains key table node n graph add node n set string label table get name string tooltip table renderer get tool tip table n set string tooltip tooltip table nodes put table n return true return false
	\end{adjustwidth}

	\vspace{0.2cm}
	\noindent\emph{\textbf{Comment}} 
	\vspace{0.1cm}
	
	\begin{adjustwidth}{0.5cm}{0.5cm}
	creates visible node for given table
    \end{adjustwidth}

    \vspace{0.2cm}
    \hrule
    \vspace{0.2cm}
    \noindent\emph{\textbf{Most similar method based on source code sequence}}

    \vspace{0.2cm}
    \noindent\textbf{Similar Method ID: 40467654}
    \vspace{0.2cm}

	\noindent\emph{\textbf{Source Code input}} 
	
	\begin{adjustwidth}{0.5cm}{0.5cm}
    public boolean is over panel int ax int ay if tab only return is over ax ay else point p new point 0 0 calc abs position p if ax p x ax p x width ay p y tab height ay p y height return true else return false
    \end{adjustwidth}

	\vspace{0.2cm}
	\noindent\emph{\textbf{Comment}} 
	
	\begin{adjustwidth}{0.5cm}{0.5cm}
	determines whether the position ax ay is over the panel takimg
	\end{adjustwidth}
	
	\vspace{0.2cm}
    \hrule
    \vspace{0.2cm}
    \noindent\emph{\textbf{Most similar method based on AST sequence}}

    \vspace{0.2cm}
    \noindent\textbf{Similar Method ID: 39298423}
    \vspace{0.2cm}

	\noindent\emph{\textbf{Source Code input}} 
	
	\begin{adjustwidth}{0.5cm}{0.5cm}
    private boolean check target reconnection line endpoints must be different shapes if new target equals old source return false return false if the line exists already for iterator iter new target get target segment part delegates iterator iter has next segment part delegate conn segment part delegate iter next return false if a old source new target line exists already and it is a differenct instance that the line field if conn get point adelegate equals old source conn equals line return false return true
    \end{adjustwidth}

	\vspace{0.2cm}
	\noindent\emph{\textbf{Comment}} 
	
	\begin{adjustwidth}{0.5cm}{0.5cm}
	return true if reconnecting the line instance to new source is allowed
    \end{adjustwidth}

\subsection{Example 2}

In this example we compare the closest methods from the source code sequence encoder and the file context encoder using the AST-Flat-FC model. This model uses a source code sequence, AST, and file context as input. The input method is a GUI method that creates a panel for cvs options. When using the source code sequence encoder to find the most similar method, we get a method that sets a view for a model. This likely was the closest method for this encoder because of tokens such as `position', `window', `grid' which all are used in many GUI methods. The file context encoder also finds a GUI method that shares language with the input method such as `minimizer' and `layout'. The file context provides additional vocabulary context, unlike the AST which learns structural similarities.

\vspace{0.2cm}
\noindent\textbf{Method ID: 299963}
\vspace{0.2cm}

	\noindent\emph{\textbf{Source code input}} 
	
	{\small
	\begin{verbatim}
private void createCVSOptions(int timeWindow){
  this.timeWindow = new TextField(
                 Integer.toString(
                      timeWindow));
  CVSOptions = new Panel();
  CVSOptions.setLayout(new GridBagLayout());
  GridBagConstraints c = new GridBagConstraints();
  c.anchor = GridBagConstraints.WEST;
  c.fill = GridBagConstraints.NONE;
  c.weightx = 1;
  c.weighty = 1;
  c.gridx = 0;
  c.gridy = 0;
  CVSOptions.add(new Label("Time window:"), c);
  c.gridx = 1;
  CVSOptions.add(this.timeWindow,c);
}
	\end{verbatim}}
	
	\noindent\emph{\textbf{Source code sequence}} 
	\vspace{0.1cm}
	{\small\begin{adjustwidth}{0.5cm}{0.5cm}
	private void create cvsoptions int time window this time window new text field integer to string time window cvsoptions new panel cvsoptions set layout new grid bag layout grid bag constraints c new grid bag constraints c anchor grid bag constraints west c fill grid bag constraints none c weightx 1 c weighty 1 c gridx 0 c gridy 0 cvsoptions add new label time window c c gridx 1 cvsoptions add this time window c
    \end{adjustwidth}}

	\vspace{0.2cm}
	\noindent\emph{\textbf{Comment}} 
	
	\begin{adjustwidth}{0.5cm}{0.5cm}
	construct the panel for the cvs options
    \end{adjustwidth}
    
    \vspace{0.2cm}
    \hrule
    \vspace{0.2cm}
    \noindent\emph{\textbf{Most similar method from source code sequence encoder}}

    \vspace{0.2cm}
    \noindent\textbf{Method ID: 45891192}
    \vspace{0.2cm}

	\noindent\emph{\textbf{Source Code input}} 
	
	\begin{adjustwidth}{0.5cm}{0.5cm}
    public void show big view state model model state view small small view m current big view new state view big model small view this m current small view small view todo move controller like state controller small is instantiate new state controller big m model this model m current big view m right panel remove all m right panel add m current big view border layout center m model set current position index of view small view refresh
    \end{adjustwidth}

	\vspace{0.2cm}
	\noindent\emph{\textbf{Comment}} 
	
	sets the model for the detailed view

	\vspace{0.2cm}
    \hrule
    \vspace{0.2cm}
    \noindent\emph{\textbf{Most similar method from file context encoder}}

    \vspace{0.2cm}
    \noindent\textbf{Method ID: 299982}
    \vspace{0.2cm}

	\noindent\emph{\textbf{Source Code input}} 
	
	\begin{adjustwidth}{0.5cm}{0.5cm}
   private void enable minimizer options boolean b dim set enabled b iter set enabled b init layout set enabled b attr exp set enabled b repu exp set enabled b grav set enabled b no weight set enabled b vert repu set enabled b load init layout set enabled b
    \end{adjustwidth}

	\vspace{0.2cm}
	\noindent\emph{\textbf{Comment}} 
	
	\begin{adjustwidth}{0.5cm}{0.5cm}
	enable the part concerning the minimizer
     \end{adjustwidth}

% \begin{figure}[t]
%     \centering
%     \includegraphics[width=0.8\columnwidth]{figures/elbow_dv.png}
%     \caption{Elbow graph for KMeans K Selection}
%     \label{fig:my_label}
% \end{figure}
\section{Discussion \& Future Work}

In this paper we present two major additions to the work in source code summarization.
\begin{enumerate}
    \item We explore the performance of ensembling a variety of baseline models using a simple aggregation technique to show how models combined with different architectures and inputs perform on the task of source code summarization.
    \item To help explain why ensembling may work well with models that use orthogonal types of input data, we explore and compare the internal source code representations these models learned. This provides insight into how we may be able to better combine models, as well as which types of models may perform best when ensembled.
\end{enumerate}

In our encoder representation comparison we discuss why complementary orthogonal input may be beneficial for models to learn better source code representations. We also show two examples to further illustrate how the learned representations of source code differs between encoders that were trained on different features of source code data (source text, AST, file context, etc). Through these examples we can see that models trained on orthogonal data complement each other well when ensembled. Also, even when training models of the same architecture, ensemble methods improve overall model performance. This paper provides a groundwork for future projects focused on the problem of ensembling source code summarization models.

\subsection{Future Work}
One potential path for future work is to explore aggregation strategies and how they can be optimized for source code summarization. In this paper we use a simple mean combination aggregation strategy and do not explore how more advanced aggregation strategies, such as meta-learning, may perform.

\section{Reproducibility}
\label{sec:reproducibility}

All of our models, source code, and data used in this work can be found in our online repository at https://bit.ly/3tiF8pc

\subsection{Hardware Details}
For training, validating, testing of our models we used a workstation with Xeon E1430v4 CPUs, 110GB RAM, a Titan RTX GPU, and a Quadro P5000 GPU

\section*{Acknowledgment}

\emph{This work is supported in part by NSF CCF-1452959 and CCF-1717607. Any opinions, findings, and conclusions expressed herein are the authors and do not necessarily reflect those of the sponsors.}
\newpage

\bibliographystyle{IEEEtran}
\bibliography{main}

\end{document}